\documentclass[useAMS,usegraphicx,usenatbib]{mn2e}

\newcommand{\barL}{\mbox{$\bar{L}$}}

\newcommand{\barM}{\mbox{$\bar{M}$}}

\newcommand{\samename}{\vrule height0.4pt depth0.0pt width1.0in \thinspace.}

\title{Near-IR Surface Brightness Fluctuations and Optical Colours of 
Magellanic Star Clusters}
\author[Gonz\'alez-L\'opezlira et.\ al]{ R.~A. Gonz\'alez-L\'opezlira$^{1}$\thanks{E-mail: r.gonzalez@astrosmo.unam.mx}
M.~Y. Albarr\'an$^{1}$, M. Mouhcine$^{2,3}$, M.~C. Liu$^{4,5}$, 
\newauthor G. Bruzual-A.$^{6}$, and B. DeBatz$^{7}$\\ \\
$^{1}$Centro de Radioastronom\'ia y Astrof\'isica, Universidad Nacional
       Aut\'onoma de M\'exico, Campus Morelia, Michoac\'an CP 58190, Mexico\\
$^{2}$School of Physics and Astronomy, University of Nottingham,
       Nottingham NG7 2RD \\
$^{3}$Observatoire Astronomique de Strasbourg (UMR 7550),
       11, rue de l'Universit\'e, 67000 Strasbourg, France \\
$^{4}$Institute for Astronomy, University of Hawaii, 2680 Woodlawn Drive, 
Honolulu, HI 96822\\ 
$^{5}$Alfred P.\ Sloan Research Fellow \\
$^{6}$Centro de Investigaciones de Astronom\'{\i}a, Apartado Postal 264, 
M\'erida 5101-A, Venezuela\\
$^{7}$GEPI (CNRS UMR 8111), Observatoire de Paris, 5 place J. Janssen, 92195 Meudon Cedex, France}

\date{Accepted ?.
      Received ?;
      in original form ?}

\pagerange{\pageref{firstpage}--\pageref{lastpage}} \pubyear{2004}

\begin{document}

\maketitle

\label{firstpage}

\begin{abstract}

This work continues our efforts to calibrate model surface 
brightness fluctuation luminosities for the study of unresolved 
stellar populations, through the comparison with data of 
Magellanic Cloud star clusters. We present here the relation 
between absolute $K_s$-band fluctuation magnitude and ($V - I$) 
integrated colour, using data from the 2MASS and DENIS 
surveys, and from the literature. 
We compare the star cluster sample with the sample of early-type 
galaxies and spiral bulges studied by \citet{liu02}. We find 
that intermediate-age to old star clusters 
lie along a linear correlation with the same 
slope, within the errors, of that defined by
the galaxies in the \barM$_{K_s}$ vs. ($V - I$) diagram.
While the calibration by \citeauthor{liu02} was determined 
in the colour range 1.05 $<$ ($V - I_c$)$_0 <$ 1.25, ours
holds in the interval -5 $\ga$ \barM$_{K_s} \ga$ -9, 0.3 
$\la$ ($V-I$) $\la$ 1.25. 
This implies, according to \citet{bruz03} and 
\citet{mouh03} models, that the star clusters 
and the latest star formation bursts in the galaxies and bulges 
constitute an age sequence. 
At the same time, a slight offset between the galaxies
and the star clusters [the latter are $\sim$ 0.7 mag 
fainter than the former at a given value of ($V - I$)], 
caused by the difference in metallicity
of roughly a factor of two, confirms that the 
\barM$_{K_s}$ vs. ($V - I$) plane may contribute to 
break the age-metallicity degeneracy in intermediate-age
and old stellar populations. The confrontation between 
models and galaxy data also suggests that 
galaxies with $K_s$ fluctuation magnitudes that are brighter 
than predicted, given their ($V - I$) colour, might
be explained in part by longer lifetimes of TP-AGB stars.
A preliminary comparison between the $H$ 2MASS data of the
Magellanic star clusters and the sample of 47 early-type galaxies and
spiral bulges observed by \citet{jens03} through the 
$F160W$ {\sl HST} filter leads to the same basic conclusions: 
galaxies and star clusters lie along correlations with the same slope,
and there is a slight offset between the star cluster 
sample and the galaxies, caused by their different metallicities.     
Magellanic star clusters are single populations, while galaxies 
are composite stellar systems; moreover, the objects 
analised live in different environments. Therefore, 
our findings mean that   
the relationship between fluctuation magnitudes in the
near-IR, and ($V - I$) might be a 
fairly robust tool for the study of stellar population ages and
metallicities, could provide additional constraints 
on star formation histories,
and aid in the calibration of near-IR SBFs for cosmological 
distance measurements.

\end{abstract}

\begin{keywords}
astronomical data bases: miscellaneous ---
galaxies: star clusters ---
galaxies: stellar content ---
infrared: galaxies --- infrared: stars ---
Magellanic Clouds ---
stars: AGB and post--AGB 
\end{keywords}

\section{Introduction} \label{intro}
Whereas surface brightness fluctuation (SBF) measurements 
are now well established as very powerful distance indicators 
\citep[e.g.,][]{tonr90,tonr97,liu01,mei01,jens01,jens03}, 
their potential for the study of unresolved stellar populations, 
although recognized long since \citep{tonr88,tonr90,liu00,blak01,
jens03}, has so far remained mostly unfulfilled. We have recently 
\citep{gonz04,gonz05,mouh05} started an effort to calibrate 
near-infrared (near-IR hereafter) SBFs, mostly for their use 
as probes of the characteristics of stellar populations. 
The idea is to compare SBFs derived from single stellar
population (SSP) models with observed fluctuation luminosities 
of Magellanic Cloud (MC) star clusters. Our contribution has 
been to build ``superclusters'' with the data, in order to reduce 
the stochastic effects produced by the inadequate representation,
in single star clusters, of the 
luminosity functions of stars evolving through short evolutionary 
phases \citep{sant97,bruz02,cerv02,cerv03,cant03,gonz04}; 
i.e., the asymptotic giant branch (AGB) and upper red 
giant branch (RGB). Superclusters, first introduced in 
\citet[][henceforth Paper I]{gonz04}, are built by coadding 
MC clusters, in the compilation of
\citet{vand81}, that have the same SWB class \citep{sear80}.
The SWB classification is based on two
reddening--free parameters, derived from integrated $ugvr$ photometry
of 61 rich star clusters in the Magellanic Clouds. 
Later, \citet{elso85} assigned SWB classes to
147 more clusters using $UBV$ photometry.
 
The SWB ranking constitutes a smooth, one--dimensional sequence that 
Searle et al.\ interpreted as one of increasing age and
decreasing metallicity; this interpretation is supported by stellar 
photometry and spectroscopy in 
individual clusters [e.g., references in \citet{sear80}; 
\citet{frog90}]. 
The sequence discovered by Searle and collaborators was arbitrarily 
segmented by them to define a classification with seven types
of clusters; however, facts like the apparently smooth progression 
in age and metallicity along the sequence, and 
the similarity between the handful of type VII MC clusters and 
the old and metal-poor halo globular clusters of the Milky Way 
have translated into the customary adoption of  
the working hypothesis that clusters within each SWB type 
have approximately the same stellar population.\footnote{
To give just a couple of examples, \citet{cohe82} 
presents ages and metallicities of the seven SWB classes
based on spectrophotometric data of a few stars within 
only a couple of clusters of each class; \citet{frog90} do the
same, based on near-IR photometry of 400 stars in and
around 39 clusters.}
In our case, as we have already explained in Paper I,
binning the data into superclusters is the compromise we have
found most convenient to adopt in order to try to circumvent
the problem of small-number statistics
posed by individual star clusters. 
We have devised eight superclusters, 
one for each of the seven different SWB classes, 
plus one ``Pre-SWB-class'' supercluster with the youngest
objects in the sample.
  
In the present paper, we focus on the relationship between \barM$_{K_s}$ 
and the ($V - I$) colour. The merits of going to infrared wavelengths 
have been discussed at length [e.g., Liu et al.\ (2000); 
Paper I; Mouhcine et al.\ (2005)]. 
Summarising, the light of intermediate-age and old populations is
dominated by bright and cool stars for which the spectral energy
distributions peak in the near-IR, at the same time that 
dust extinction is reduced. On the other hand, there is a scarcity 
in the near-IR of both accurate empirical calibrations and 
self-consistent models, a fact that we hope we are contributing 
to remedy with our work.

In the particular case of the correlation  between \barM$_{K_s}$ 
and ($V - I$), 
\citet{liu02} confirmed  
a linear dependence of \barM$_{K_s}$ with ($V - I$) in a sample 
of 26 ellipticals, S0s, and spiral bulges in the Local Group, 
Fornax, Virgo, Eridanus, and Leo. The relation had already 
been hinted at by the data set of \citet{jens98}, which covered
only half the range in colour [$\sim 1.16 <$ ($V - I$) 
$< \sim$ 1.24]; it was more clearly shown 
by \citet{mei01}, mainly through the inclusion of data 
for NGC 4489 at ($V - I$) $\la$ 1.05. 
\citeauthor{liu02} filled significantly the gap in colour 
with observations of Fornax objects.
Probing the extent (in age, 
metallicity and, ultimately, environment) to which this relation 
remains valid has obvious implications for an accurate calibration 
of near-IR SBFs as distance indicators, and hence for a precise 
determination of $H_{\circ}$ through these measurements [see 
\citet{liu02} for a detailed discussion].

There are also implications of this correlation from the
point of view of stellar population studies.
Since in general observed properties of stellar populations 
are luminosity-weighted, and given that
after $\sim$ 12 Myr the youngest stars are 
also the most luminous in the optical and near-IR wavelengths, 
the tendency uncovered by \citet{liu02} could be tracing 
the most recent burst of star formation in each of these systems.
This is the conclusion reached in the cited work, after comparing
their results with predictions by \citet{liu00} [based on 
the models that would later be published by \citet{bruz03}],
\citet{wort93b}, and \citet{blak01} [based on the \citet{vazd96} 
models].  Although there are disagreements among such predictions 
(especially in the derived metallicities obtained, respectively, 
from the \citeauthor{liu00} models on one hand, and from those 
of \citeauthor{blak01} on the other), all of them point toward
a large spread in age for the sample, from less than 5 Gyr to more 
than 12 Gyr. Furthermore, in all three sets of models
confronted by \citet{liu02}, the age and metallicity sequences 
are not parallel in at least some regions of the 
\barM$_{K_s}$ vs. ($V - I$) plane, opening the possibility 
of using it as a diagnostic for breaking the age-metallicity degeneracy.

This study of the relationship between \barM$_{K_s}$ and 
($V - I$) in the MC star clusters is complementary in two advantageous 
ways to the investigation by \citet{liu02}. Firstly, since the 
superclusters are approximately single age, single metallicity stellar
populations,
the star formation bursts that they represent are not masked by 
an underlying population, as it occurs in galaxies; hence, they likely 
constitute a better set for comparison with simple stellar 
population models. Secondly, the MC clusters span an even larger extent 
in age, i.e., from a few $\times 10^6$ yr to $\sim 10^{10}$ yr, than that 
of the galaxies and bulges in the \citeauthor{liu02} sample. 
Consequently, they cover a range roughly three times larger in 
\barM$_{K_s}$ and four times more extended in ($V - I$).


For this investigation, we have made use of $K_s$-band data 
retrieved from the Two Micron All Sky Survey \citep[2MASS;][]{skru97}; 
$I_{Gunn}$ ($I_g$ or $I$
henceforth) data from the Deep Near-Infrared Southern Sky Survey    
\citep[DENIS;][]{epch97}; and $V$ data from different sources in
the literature.
This paper is organized as follows. In \S~\ref{data}, we give a 
brief summary of the data acquisition and characteristics, as well 
as of our own treatment of such data. In \S~\ref{theo_prd}, we
present the ingredients of the stellar population synthesis models,
and a comparison between the two sets of models used here. 
In \S~\ref{results}, we compare the theoretical predictions to 
the observations. Finally, in \S~\ref{concl}, the results of the
present work are discussed and summarised.

\section{Observational data of Magellanic Cloud star clusters} \label{data}
\subsection{The 2MASS data}
\label{2mass}

$K_s$ data for 191 MC clusters from the compilation of
\citet{vand81} and analised by Elson
\& Fall (1985, 1988) were retrieved from the 2MASS archive.
The data processing has been presented at length in 
Paper I. Succinctly, the fluctuation luminosity is the ratio
of the second moment of the luminosity function
($\Sigma n_i L_i^2$) to its first moment (the integrated
luminosity, $\Sigma n_i L_i$), as expressed by the
equation:

\begin{equation}
\barL \equiv \frac{\Sigma n_i L_i^2}{\Sigma n_i L_i}.
\label{lbareq}
\end{equation}

\noindent
Bright stars are the main contributors to the numerator, whereas
faint stars contribute significantly to the denominator.
The second moment of the luminosity function 
was derived by measuring the flux of resolved, bright stars in the MC
clusters, 
while the integrated luminosity was computed from the total light 
detected in the images, after removing the sky background emission. 
In order to reduce the stochastic errors produced by the small 
numbers of luminous and cool RGB and AGB stars in single star 
clusters, eight superclusters were assembled with the 2MASS 
data, one for each of the seven different SWB classes, 
plus one ``pre-SWB-class'' supercluster; this was accomplished 
by stacking individual clusters with the same SWB type.\footnote{
Individual cluster images were multiplicatively scaled to
a common photometric zero-point (determined
by the 2MASS team) 
and dereddened; SMC clusters
were geometrically magnified to place them at the distance of the
LMC.} 
The mosaics were used to measure the 
integrated light of the superclusters. To derive the second moment 
of the luminosity function, star lists for each one of the 
superclusters were integrated with entries from the 2MASS Point 
Source Catalog (PSC). The photometry of the point sources 
was performed by the 2MASS collaboration on individual
cluster frames, previously to and independently from this work 
(i.e., we did not measure 
fluxes from point sources on the stacked supercluster frames), 
following the standard procedure of profile-fitting plus 
curve-of-growth aperture correction.\footnote{
http://www.ipac.caltech.edu/2mass/releases/allsky/doc/sec4\_4.\\html}
Since all near-IR zero-points were obtained 
in a uniform fashion, it is unlikely that they are 
an important source of systematic error.   

As explained in Paper I, the quoted fluctuation errors include
stochastic variations produced by small-number statistics. 
These were calculated following a statistical approach 
introduced by \citet{buzz89} and \citet{cerv02}, 
based on the assumption that
the variables involved have a Poissonian nature. 
In this framework, stochastic errors scale as 
$M_{tot}^{-1/2}$, where $M_{tot}$ is the total
mass of the stellar population. 
We refer the reader to the appendix for a short discussion 
on the subject of stochastic variations.

Besides the problem of small-number statistics, 
assessing crowding is 
crucial for accurate SBF measurements of star clusters. Through 
the blending of sources, crowding can in principle make
the numerator of equation \ref{lbareq} larger and 
hence the SBF magnitude brighter. Another source of systematic
error is the sky level, which impacts the denominator of
equation \ref{lbareq}. Both crowding and sky determination 
have been lengthily addressed in Paper I, and revisited in 
\citet{mouh05} to consider the corrected fluctuation measurements 
of superclusters type I and II \citep{gonz05}. 
In regard to crowding, we did not follow there
the usual procedure of analising the effects of the addition
of artificial stars on the measured fluctuations. Instead, we
compared the fluctuations derived from annular regions  
of the superclusters (i.e., regions with diverse crowding properties),
and checked whether the different regions would be deemed crowded 
by the criterion, developed by \citet{ajha94}, that the two
brightest magnitudes of stars cover more than 2\% of the area.
We also tested our hypothesis that the PSC quality flags would 
be very helpful to eliminate blended sources by obtaining, again
in annuli, fluctuation values with and without stars 
with faulty photometry.
While crowding affects preferentially the centers of clusters,
errors in the sky subtraction will 
impact more the fainter, outer regions. 
Our analysis allowed us to determine that  
the circular regions within
1\arcmin\ of the center of the superclusters provide the
better balance of uncertainties owing to crowding and
sky subtraction, at the same time that they are less vulnerable
to stochastic effects than smaller annular regions. 
Consequently, for the SBF measurements  
we use only integrated light and point sources within 1\arcmin\ 
(at the distance of the LMC) from the centres of the superclusters. 
Just point sources with good photometry, according to the PSC quality 
flags,\footnote{
http://www.ipac.caltech.edu/2mass/releases/allsky/doc/sec1\_6b.\\html\#origphot}
were included. We adopted distance moduli of 
$(m - M)_{\rm o} = 18.50 \pm 0.13$ to the LMC and 
$(m - M)_{\rm o} = 18.99 \pm 0.05$ for the SMC \citep{ferr00}.

\subsection{The DENIS data}
\label{denis}

The observations for DENIS 
were carried out between 1995 and 2001, and in particular 
the MC data were taken between 1995 and 1998. \citet{cion00} give 
a good summary of the DENIS instrument, and data acquisition and
characteristics. The instrument was mounted
at the Cassegrain focus of the 1-m ESO telescope at La Silla, Chile,
and could obtain simultaneously images at $I_g$,  
$J$, and $K_s$ with 3 cameras. These had, respectively,
a Tektronix CCD with 1024$^2$ pixels 
(each 1\arcsec$\times$1\arcsec), and two NICMOS infrared detectors 
with 256$^2$ pixels (each 3\arcsec$\times$3\arcsec, but the
$J$ and $K_s$ exposures were dithered to a 1\arcsec pseudo-resolution).
DENIS scanned the Southern sky in strips of 30\degr\ in declination 
and 12\arcmin\ in right ascension. Each strip comprises 
180 12\arcmin$\times$12\arcmin\ images, with an overlap of 2\arcmin\ 
between every pair. The integration time at $I$ is 9 s; 
at $J$ and $K_s$, the integration time is 1 s, but each 
released image is made up of 9 individual microscanned 
exposures, for a total integration time also of 9 s.  

The nominal 5$\sigma$ limiting magnitude at $I$ is 18 mag,  
and the typical size of a detected point source is smaller than 
2\arcsec\ FWHM. However, for this particular project we are 
concerned with the photometric accuracy at $I$ achieved for 
extended sources. From those clusters for which more than one $I$ 
calibrated exposure is available, we find the photometric 
error to be $\sim$ 0.1 mag.

Flattened and bias-subtracted $I$ images that contain the 
Magellanic clusters in our sample were retrieved from the DENIS 
archive. The sample is the same one from which the \barM$_{K_s}$ 
measurements were obtained, with the exception of seven clusters, 
presented in Table 1; there are no calibrated DENIS 
data for six of them, while the other one (NGC~1777) has two bright 
foreground stars and we do not know whether the $V$ measurement 
has been duly corrected. Instrumental zero-points were obtained 
from the DENIS archive as well. Next, the $I$-band flux was derived 
for each cluster, in a diaphragm with the same size as the one used 
for the corresponding $V$-band measurement. $V$ magnitudes and 
diaphragms were taken from \citet{vand81}, except for those clusters 
listed in Table 2.\footnote{In the case of the 2 clusters 
for which only  $F555W$ radial profiles \citep{mack03} were available, 
integrated magnitudes were derived with diaphragms of 60\arcsec.} 
It is worth noticing that, since we do not know the coordinates 
of the centres of the $V$ observations, the derived ($V - I$) colour 
could be slightly biased to the red.\footnote{When we change the 
input centre for the $I$ photometry  --we do not have $V$ images-- 
between 5 and 10\arcsec, the ($V - I$) colour 
of a single cluster shifts by about 0.1 mag.} The sky emission for each cluster 
was determined from an annulus separated from the photometric diaphragm 
by a buffer area; the best sizes of both buffer and sky annuli were
chosen after visually inspecting the images, with the purpose of 
excluding from the sky measurements bright foreground 
stars and, most importantly, residual cluster light. 

The individual cluster $V$ and $I$-band flux values were then corrected 
for extinction as in Paper I, Table 1, and averaged 
to obtain the $V$ and $I$ fluxes for the superclusters, and their ($V - I$) 
colours. 
As before, when clusters 
do not have individually measured reddening, we have 
assumed $E(B-V) = 0.075$ for the LMC and $E(B-V)=0.037$ for the SMC 
\citep{schl98}.
Once again, given that the extinction corrections of all
the data ($V$, $I$, and $K_s$) were done consistently by us, 
this is probably not an important source of systematic error.
Absolute zero points were taken, for $V$ from 
\citet{bess79}, and for $I$ from \citet{fouq00}, who
determined it especially for DENIS. The average ($V - I$) 
colours of the superclusters [and, for completeness, also the 
$\barM_{K_s}$ values derived in Paper I and in \citet{gonz05}] 
are presented in Table 3. 
The quoted errors in ($V - I$) were derived as follows: the dispersion 
of $I$ flux measurements was calculated for those clusters for which 
multiple images were available; for clusters with single images, 
the average dispersion was adopted; the error in $I$ for the 
superclusters was found by adding the individual dispersions in 
quadrature, and then dividing by $(N - 1)^{1/2}$, with $N$ the 
number of individual clusters in each supercluster; finally, 
the error in ($V - I$) was computed by assigning 
to the $V$ value the same uncertainty as the one for $I$, and assuming 
a correlation coefficient of 0.5 between $V$ and $I$ [see, for example, 
\citet{cerv02}; Paper I].\footnote{The compilation by \citet{vand81} does 
not quote errors for the $V$ photometry, but marks as uncertain  
($U - B$) or ($B - V$) values for which any two observations 
differ by more than 0.1 mag. Among van den Bergh's sources, 
only \citet{vand68} and \citet{alca78} list errors for $V$. 
Hence, the clusters in our work with published errors go from 
18\% for SWB class VI to 75\% for SWB type VII. For supercluster 
type VII, we have calculated the uncertainty in ($V -I$) using the 
errors in these papers; it differs only 
by 0.01 mag from the one derived with the procedure described above.}
We note, however, that while we find an 
average dispersion at $I$ of $\sim$ 0.1 mag for individual clusters, 
there might be systematics, like flattening defects, that would 
increase this error somewhat. Indeed, \citet{patu03} give an average
uncertainty of $\sim$ 0.2 mag at $I$ for their measurements of galaxy
magnitudes. We also remark that the dispersion of ($V - I$) colours 
among the individual constituents of each supercluster is typically
$\sim$ 0.2 -- 0.3 mag. Although some fraction of this spread could 
be attributed to the fact that superclusters are not really 
single stellar populations, it is nevertheless consistent with
the scatter among different low-mass realizations of a true SSP.
For example, using Monte Carlo simulations, \citet{bruz02} 
finds 3-$\sigma$ fluctuations in ($V - K$) of almost 2 mag for
a 1$\times 10^4 M_\odot$ cluster at a given age.
Moreover, the distribution of
($V - I$) colours of individual clusters within each supercluster
looks reasonably Gaussian for all except supercluster 
SWB VII. There are only 12 star clusters type VII, and while
the mode of the distribution (3 clusters) lies around 
($V - I$) = 1, there is a tail (2 clusters) 
with ($V - I$) $>$ 1.4. This tail could be produced by 
real population differences or, again, by small-number statistics. 


\section{Stellar population model predictions}
\label{theo_prd}

Central to this paper is the comparison between, on the one hand, 
the observed optical photometry and $K_s$-band fluctuation magnitudes 
of MC superclusters, and, on the other, the properties of synthetic 
single age, single metallicity stellar populations as predicted by, 
respectively, Bruzual \& Charlot (2003, BC03) and Mouhcine \& Lan\c{c}on 
(2003, ML03) stellar population synthesis models. BC03 models range 
from 0.1~Myr to 17~Gyr in age, while those of ML03 go from 12~Myr to 
16~Gyr. Both sets of models span initial stellar metallicities of
$Z/Z_{\odot}=1/50$ to $Z/Z_{\odot}=2.5$. Here, we 
describe briefly the main ingredients of the stellar population 
synthesis models, referring to the quoted papers for more details.
Note that other sets of theoretical predictions of near-IR $K_s$-band
surface brightness fluctuation magnitudes and optical colours are
published; however, they generally use Bruzual \& Charlot (2003) 
isochrones or one of their earlier versions (Liu et al. 2001; 
Mei et al. 2001), or they are computed for old stellar populations 
only, i.e., they do not cover the stellar population age range 
spanned by the MC superclusters (e.g., Worthey 1993a,b; Blakeslee 
et al. 2000; Cantiello et al. 2003).

\subsection{Single Stellar Population Models}

ML03 stellar population synthesis models
were designed to reproduce the near-IR properties of both resolved 
and unresolved stellar populations, with an emphasis on intermediate 
age stellar populations. The library of evolutionary tracks used  
by ML03 is based on the models of Bressan 
et al. (1993) and Fagotto et al. (1994\,a,b,c). We will refer to 
these sets as the Padova tracks hereafter. 
The sets of tracks cover major stellar evolutionary phases: from 
the main sequence to the end of the early-AGB phase for low- and 
intermediate-mass stars, and to the central carbon ignition for 
massive stars. The Padova tracks do not extend to the end of the 
Thermally Pulsing Asymptotic Giant Branch phase (TP-AGB hereafter). 
The high luminosities and low effective temperatures of stars
evolving through this phase make them among the main 
contributors to the integrated near-IR light of stellar systems within 
the age interval when these stars are alive (e.g., Frogel et al. 
1990; Mouhcine \& Lan\c{c}on 2002). 
The extension of these tracks to cover the TP-AGB phase is then 
needed. Until this happens, the evolution of low- and intermediate-mass 
stars through the 
TP-AGB evolutionary phase is followed using the so-called synthetic 
evolution modelling (e.g., Iben \& Truran 1978; Renzini \& Voli 
1981; see also Marigo et al. 2003 for another attempt to include 
the TP-AGB in stellar population models). The complex interplay between 
different processes affecting stellar evolution through the TP-AGB 
phase are taken into account in the synthetic evolution models 
used by ML03. 
The properties of TP-AGB stars are allowed to evolve according 
to semi-analytical prescriptions. These prescriptions take into 
account the effect of metallicity on the instantaneous properties 
of TP-AGB stars. The evolution of TP-AGB stars is stopped at 
the end of the AGB phase, since their contribution to the 
optical/near-IR light is almost negligible once they evolve 
beyond this phase. The models predict the effective temperatures, 
bolometric luminosities, lifetimes, and relative lifetimes of the 
carbon-rich phase of TP-AGB stars.

The BC03 models we will compare to the observed properties of MC 
superclusters are those that these authors call the {\it standard 
reference models}. These models were built using isochrones based 
on Padova stellar evolutionary tracks. To extend the Padova low- 
and intermediate-mass stars beyond the early-AGB, BC03 adopt the 
effective temperature, bolometric luminosities, and lifetimes of 
TP-AGB stars from Vassiliadis \& Wood (1993). The relative 
lifetimes in the carbon-rich phase for TP-AGB stars were taken 
from Groenewegen \& de Jong (1993), and Groenewegen et al. (1995).

Stellar libraries are needed to assign spectral energy distributions
to the stars of a synthetic population. For stars evolving through
phases other than the TP-AGB, ML03 have used 
the theoretical stellar atmospheres of Kurucz (see Kurucz, 1979), 
Fluks et al. (1994) and Bessell et al. (1989, 1991), as collected 
and re-calibrated by Lejeune et al. (1997, 1998). BC03 
use also the atmosphere compilations by 
Lejeune et al., as corrected 
by Westera et al. (2001, 2002).
These stellar spectral libraries do not include spectra of TP-AGB 
stars. The spectrophotometric properties of TP-AGB stars are different 
from those of other cool and luminous stars. Their extended and cool 
atmospheres lead to the formation of deep and specific spectral 
absorption bands. On the other hand, the photometric properties of  
oxygen-rich and carbon-rich TP-AGB stars are different, i.e., as an 
example, carbon stars show redder ($H - K$) colours than oxygen-rich TP-AGB 
stars at a given ($J - H$) colour. To account for TP-AGB star properties, 
ML03 have used the empirical library of average TP-AGB 
oxygen-rich and carbon-rich star spectra of Lan\c{c}on \& Mouhcine 
(2002). The empirical average spectra of carbon stars
from Lan\c{c}on \& Mouhcine (2002) are used at all metallicities.
It is extremely difficult, however, to estimate the effective 
temperature and the metallicity of TP-AGB oxygen-rich stars; 
instead, ML03 use a metallicity-dependent effective temperature scale 
(Bessell et al. 1991). 
To account for the spectral properties of carbon-rich TP-AGB stars, 
BC03 have constructed period-averaged spectra using solar 
metallicity model atmospheres for carbon stars from H\"ofner et al. 
(2000). These spectra were used to represent carbon stars of all 
metallicities in BC03 models.

$K_s$-band fluctuation magnitudes were computed from the models, 
as has been amply described in Paper I and summarised in 
\citet{mouh05}. In this case, the second moment 
of the luminosity function is calculated by summing $L_{K_s}^2$ over all stellar 
types in the models; the integrated luminosity is obtained by summing the 
$L_{K_s}$ of all types of star. Correspondingly, model 
($V - I$) colours were constructed by summing the $V$ and $I$ fluxes 
of all the stars in the isochrones.  

\subsection{Comparison between models} \label{compmod}

\begin{figure*}
\includegraphics[clip=,width=0.32\textwidth]{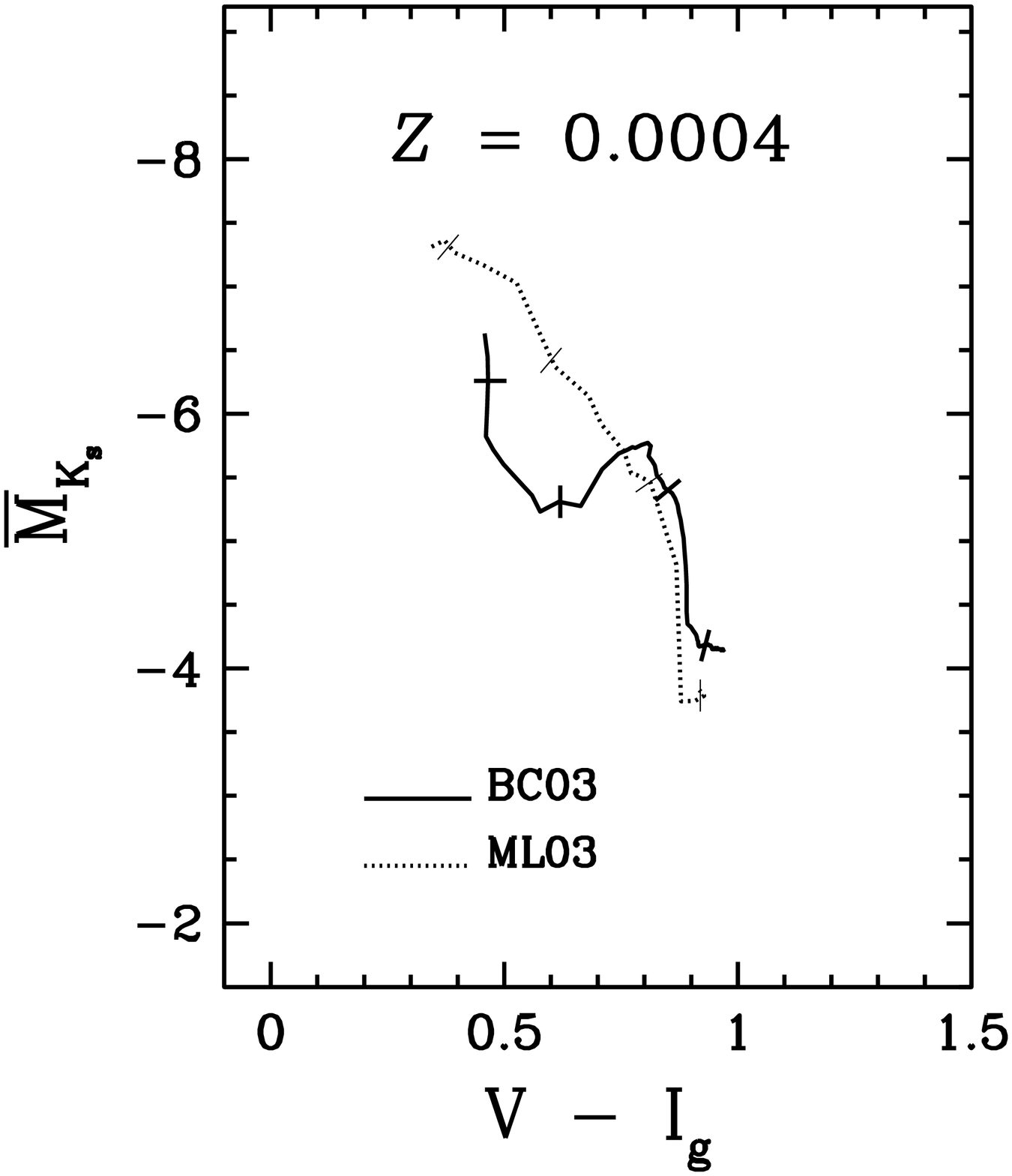}
\includegraphics[clip=,width=0.32\textwidth]{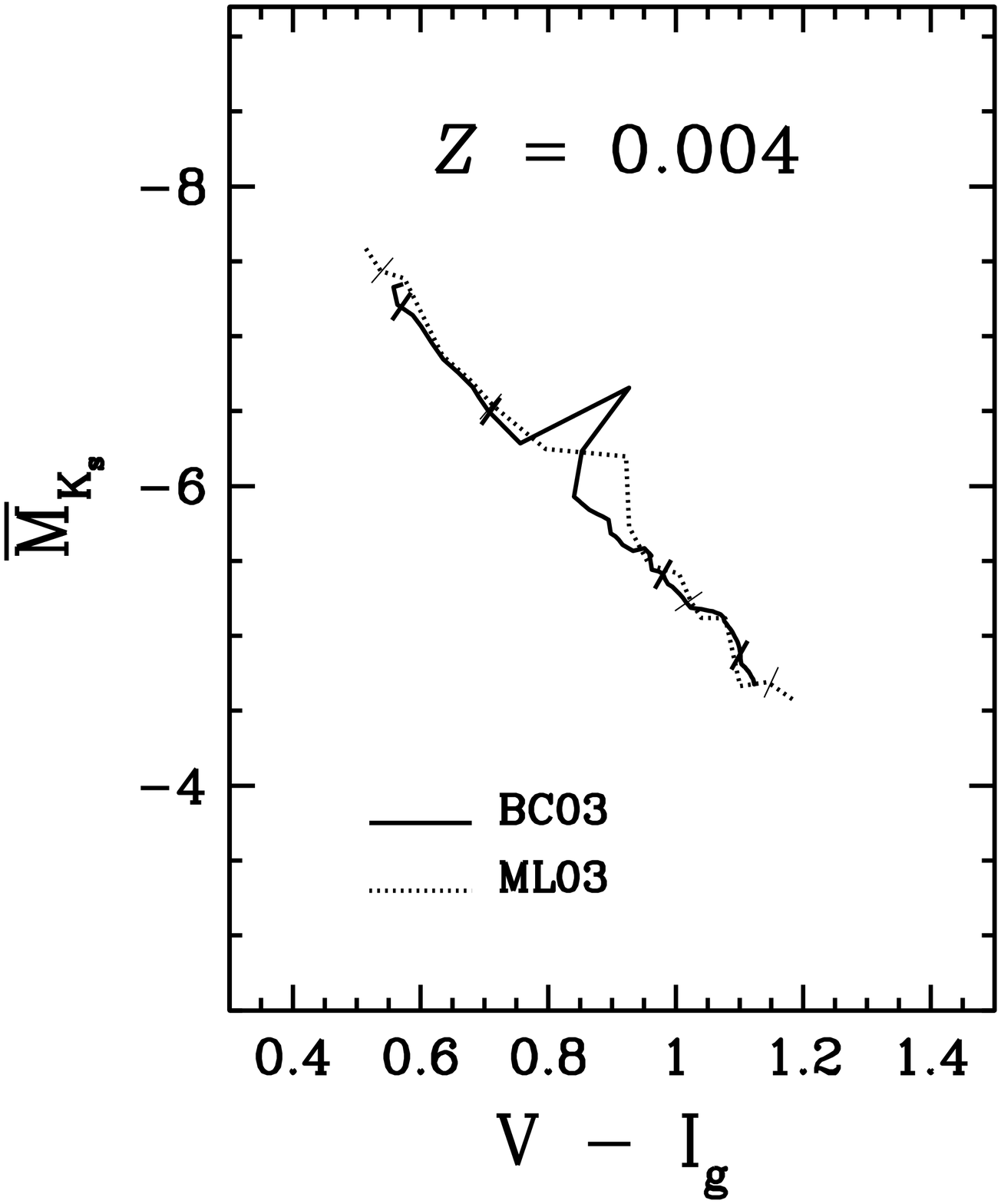}
\includegraphics[clip=,width=0.32\textwidth]{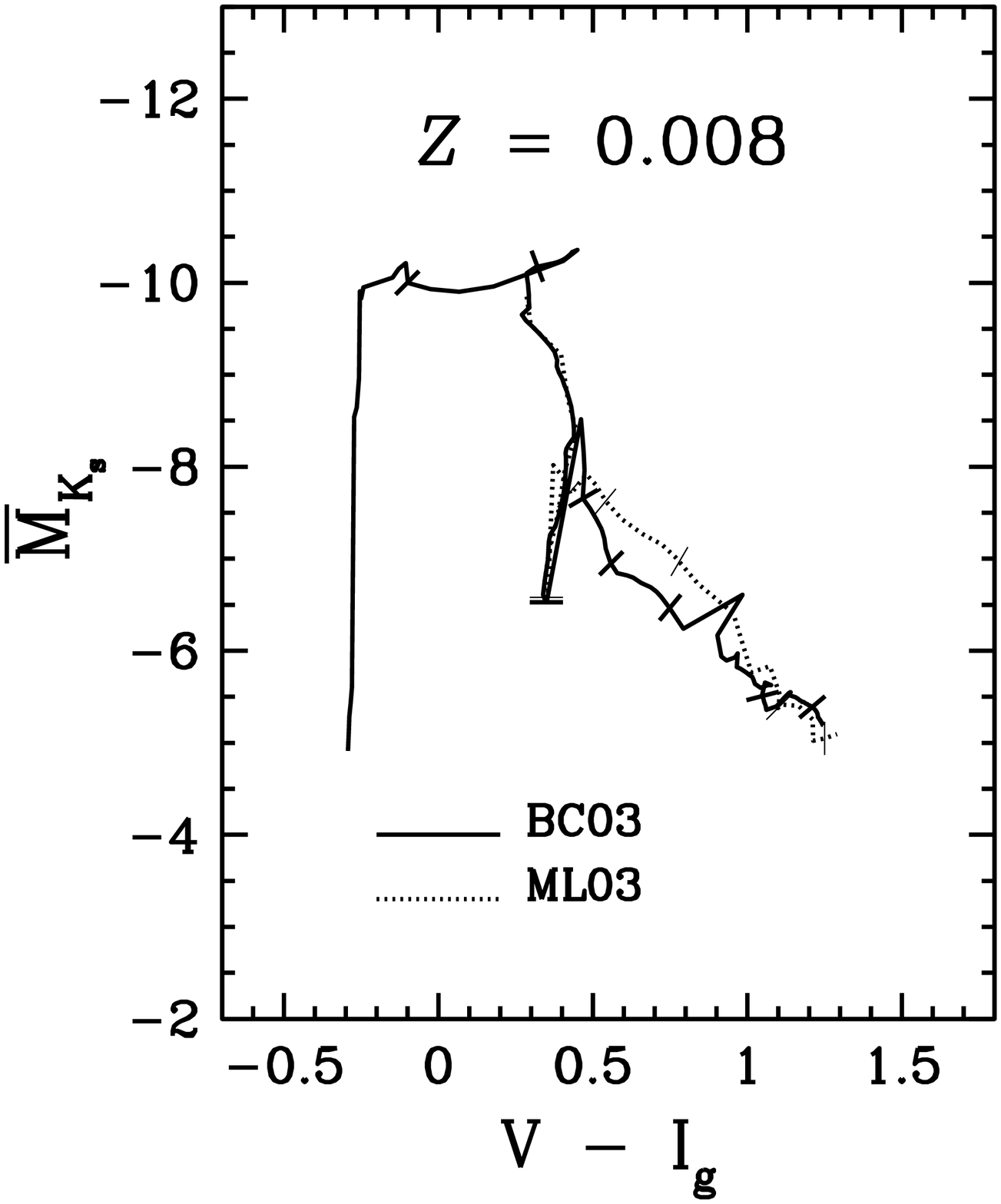}
\includegraphics[clip=,width=0.32\textwidth]{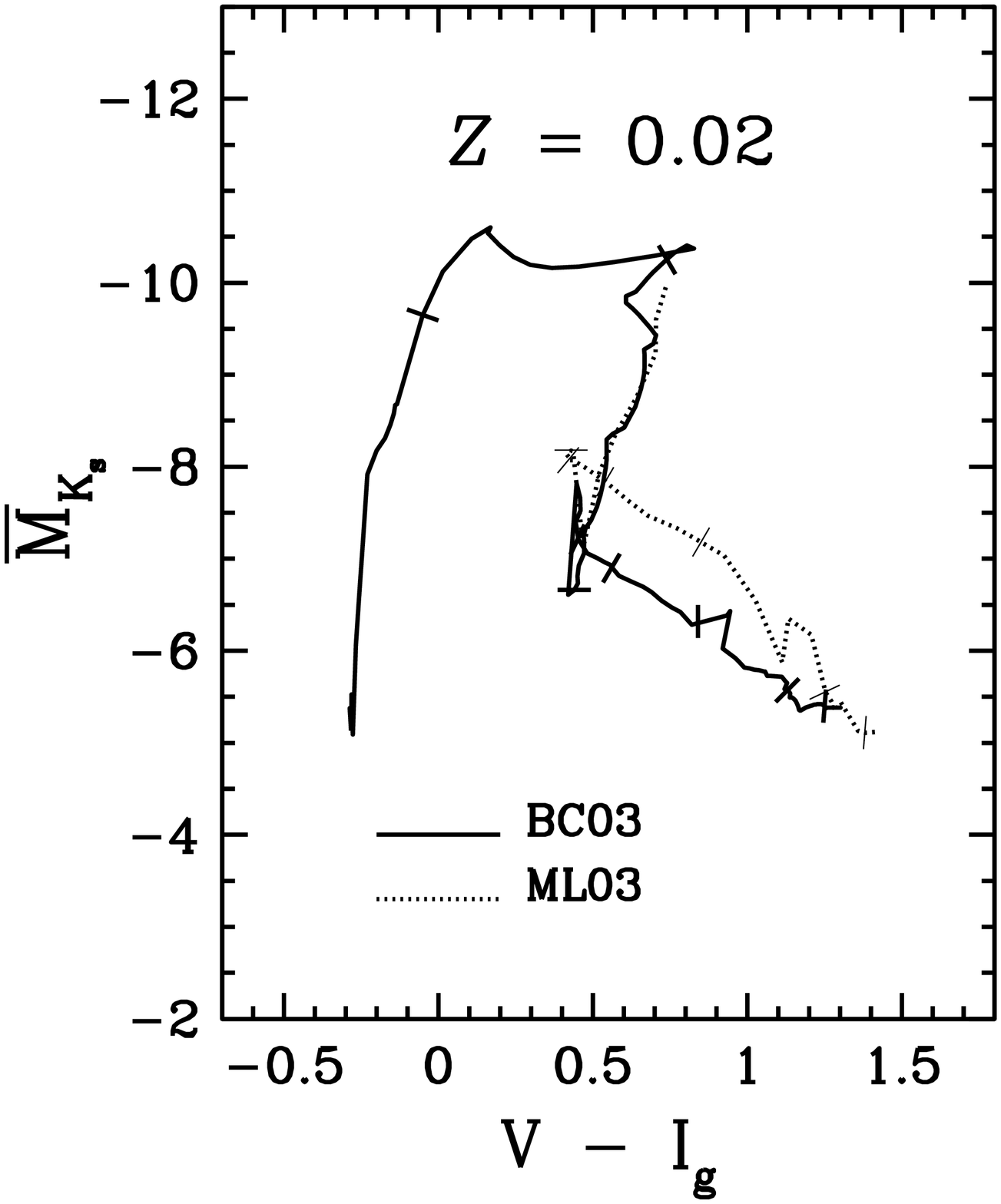}
\includegraphics[clip=,width=0.32\textwidth]{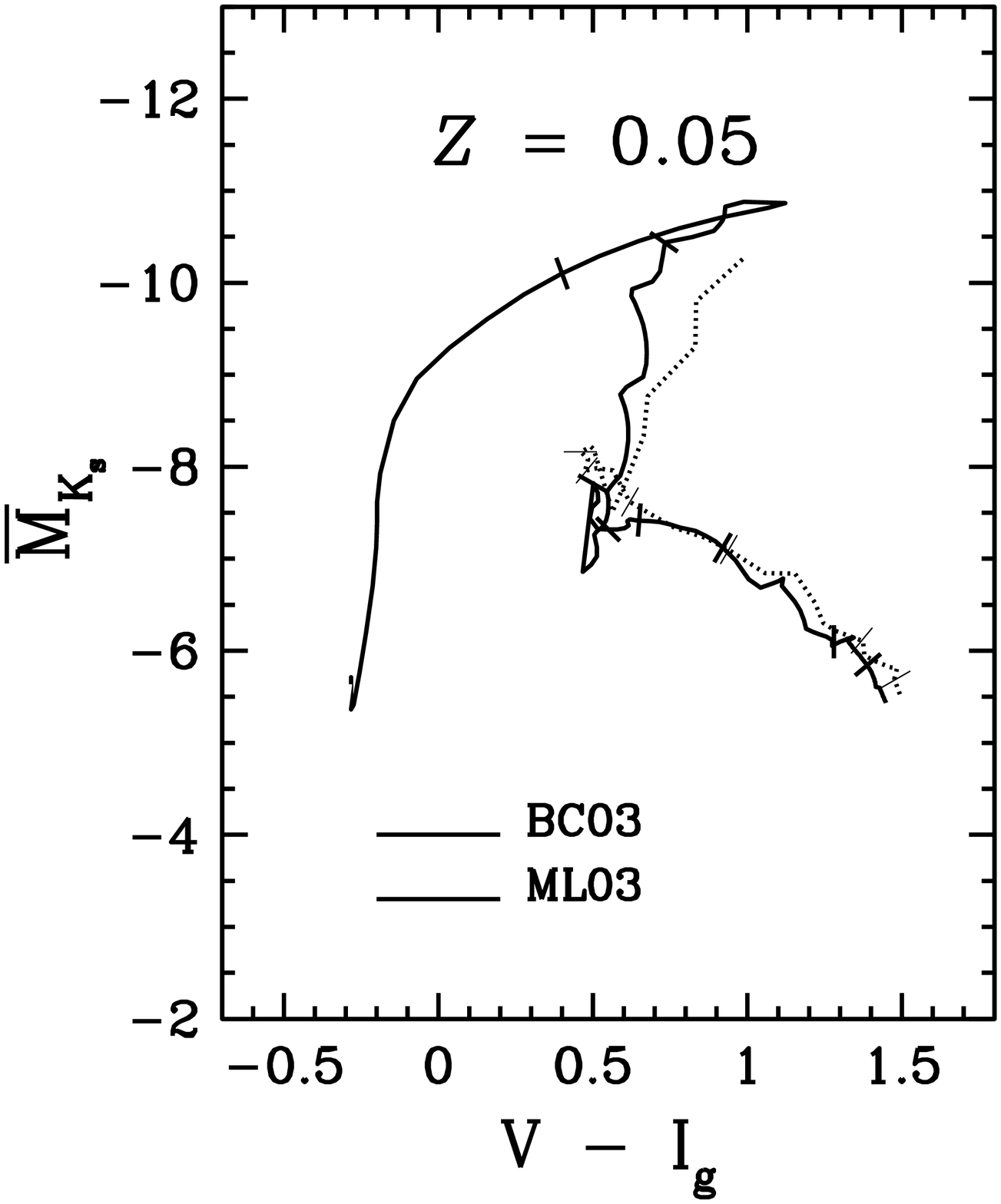}
  \caption{
Comparison of model \barM$_{K_s}$ versus ($V - I_g$) colour 
for $Z$ = 0.0004 ({\it top left}), $Z$ = 0.004 ({\it top middle}),
$Z$ = 0.008 ({\it top right}), $Z$ = 0.02 ({\it bottom left}), 
and $Z$ = 0.05 ({\it bottom middle}). {\it Solid lines:} 
Bruzual \& Charlot (2003, BC03); {\it dotted lines:} 
Mouhcine \& Lan\c{c}on (2003, ML03). For $Z$ = 0.0004 and 
$Z$ = 0.004, evolution of models is shown only between 
$\sim$ 300 Myr and $\sim$ 16 Gyr. In the remaining panels, 
BC03 and ML03 models start, respectively, at 2.4 Myr and 12 Myr.
From left to right, tickmarks indicate 5 and 10 Myr (only for BC03,  $Z \ga$ 0.008 
models), 100 and 200 Myr (for BC03 and ML03 models with 
$Z \ga$ 0.008), 0.4, 1, 4, and 12 Gyr (all models).}
\label{mdl_comp}
\end{figure*}

Fig. \ref{mdl_comp} presents a comparison between the evolution 
of single-age, single-metallicity stellar population properties 
in the \barM$_{K_s}$ vs. ($V-I$) diagram as predicted by BC03
and ML03 stellar population synthesis models, respectively, for 
all the metallicities considered. The figure shows that both 
sets of models predict a qualitatively similar evolution over 
the age range comprised by them; i.e., the $K_s$-band surface 
brightness fluctuation magnitudes get fainter as the ($V-I$) 
colour gets redder. For both sets of stellar population model 
predictions, the ($V-I$) colour increases gradually to redder 
values as the stellar populations age. On the other hand, it 
is predicted by both BC03 and ML03 models that at a fixed age, 
single age, single metallicity stellar populations show redder 
($V-I$) colours at higher metallicity. This is because, at
fixed initial stellar mass, lowering metallicity causes stars
to evolve at higher effective temperatures.
In view of the monotonic and smooth evolution of the ($V-I$)
colour, and given its weaker sensitivity (compared to 
the near-IR wavelength range) to the presence of 
very cool and luminous stars, this colour index can be 
regarded as a primary age indicator at fixed metallicity. 

For the stellar populations dominated by red supergiant stars, 
i.e., younger than a few $\times$ 10\,Myr, the $K_s$-band 
SBF magnitude gets drastically fainter with age; 
on the other hand, the ($V-I$) changes by a modest
factor around $\sim$ 0.3 mag. 
This trend is produced by the combination of two 
facts: (1) the masses, and hence the luminosities, of the 
red supergiant stars that drive the 
fluctuation signal change significantly, while (2) the 
combined mass of main sequence stars, that determine the 
integrated optical properties, stays almost
constant.

When red supergiant stars disappear from a stellar population, 
AGB stars drive the evolution of the 
near-IR properties up to 1.5--2 Gyr. For 
populations dominated by short-lived massive AGB stars, i.e., 
younger than ${\rm \sim\,200~Myr}$, the predicted evolution 
in the \barM$_{K_s}$ vs. ($V-I$) 
diagram is complex. When the first (massive) AGB stars emerge 
in the stellar population, a brightening of the $K_s$-band surface 
brightness fluctuation magnitude is predicted, at almost fixed 
($V-I$) colour. This is because of the overluminosity produced 
by the envelope burning that affects AGB stars with large initial 
stellar masses, i.e., ${\rm M_{init}\ga\,3.5-4\,M_{\odot}}$
(see e.g., Mouhcine \& Lan\c{c}on 2002 for more details on the
effects of envelope burning on intermediate-age stellar
population properties). 
The observed counterparts of single age, single 
metallicity stellar populations within this age range are 
expected to cluster at the same location in the \barM$_{K_s}$ 
vs. ($V-I$) diagram, i.e., in the region around $(V-I)\sim\,0.4$ 
and $\barM_{K_s}\sim\,-7.5$. 
During this age interval, the models have 
no ability for stellar population age-dating, and/or metallicity 
estimate. 
 
For stellar populations older than ${\rm \sim\,300~Myr}$, in 
which AGB stars are not affected by envelope burning, both 
sets of models predict a monotonic dimming of 
the $K_s$-band surface brightness fluctuation magnitudes as the 
($V-I$) colour gets redder. This evolutionary pattern continues
for older stellar populations, i.e., older than 2--3 Gyr, when
RGB stars drive their near-IR properties. 
This is because of the evolution of late-type giant star content. 
As a stellar population ages, the mass of the stars fueling 
the evolution of near-IR properties, i.e., AGB stars for ages 
younger than $\sim\,1.5\,$Gyr, and red giant stars for older 
ages, decreases. Consequently, at a fixed metallicity, the average luminosity 
of these stars decreases with stellar 
population age. 
Conversely, at a fixed age, \barM$_{K_s}$ magnitudes 
get brighter and the ($V-I$) colour gets redder
as the stellar population metallicity increases.
Thus, both the BC03 and ML03 sets of models 
predict that, as the metallicity increases, 
populations with the same age move  
to the upper right in the \barM$_{K_s}$ vs. ($V-I$) 
diagram.

Despite the qualitative agreement between the BC03 and ML03 sets
of theoretical predictions, differences between the two are 
apparent. At a given stellar metallicity, 
the evolutionary track 
in the \barM$_{K_s}$ vs. ($V-I$) diagram predicted by ML03 
models is systematically redder than the one 
predicted by BC03 models. For stellar populations with 
$Z$= 0.0004, 0.008, and 0.02, and between ages 300~Myr 
and 1.5~Gyr, the models based on ML03 isochrones predict, at 
similar ($V-I$) colours, brighter $K_s$-band SBF magnitudes 
[see also \citet{mouh05}]. Observationally, this 
means that for a given ($V-I$) colour, 
models based on BC03 isochrones will attribute 
a higher metallicity to a star cluster with a certain 
\barM$_{K_s}$ magnitude. The $K_s$-band SBF 
magnitude in the models based on ML03 isochrones is more 
sensitive to the presence of AGB stars. 
This is due to the fact 
that the AGB lifetimes used in the ML03 stellar population synthesis 
models are larger, thus increasing 
the contribution of these stars to the $K_s$-band light budget.  

\begin{table*}
 \centering
 \begin{minipage}{140mm}
  \caption{Clusters in Gonz\'alez et al.\ (2004) not included 
in this paper}
  \begin{tabular}{@{}lcl@{}}
  \hline
  Supercluster&Name&Reason \\
 \hline
I\dotfill & L~51 & No available calibrated DENIS data \\
& SL~477 & No available calibrated DENIS data \\
& NGC~1951 & No available calibrated DENIS data \\
& NGC~1986 & No available calibrated DENIS data \\
III\dotfill& NGC~1953 &  No available calibrated DENIS data\\
V\dotfill&  NGC~1777 & Two bright foreground stars\\
&  NGC~2193 & No available calibrated DENIS data\\
\hline	
\end{tabular}
\end{minipage}
\label{tabmiss}
\end{table*}

\begin{table*}
 \centering
 \begin{minipage}{140mm}
  \caption{$V$ photometry sources different from \citet{vand81}}
  \begin{tabular}{@{}lcl@{}}
  \hline
  Supercluster&Name&Source \\
 \hline
Pre-SWB \dotfill& NGC~1727 & \citet{bica96} \\  
& NGC~1936 (IC~2127) &  \citet{bica96} \\
& NGC~2014 &  \citet{bica96} \\
& NGC~2074 &  \citet{bica96} \\
& NGC~2001 &  \citet{bica96} \\
I\dotfill&      NGC~299 & \citet{alca78}\\
IV\dotfill& SL~663 & \citet{mack03}\footnote{
$F555W$ magnitudes were first obtained by 
integrating over the radial luminosity profiles published 
by \citet{mack03}, then transformed to $V$ with the
relations of \citet{dolp00} (without 
charge transfer terms), assuming ($B - V$) colours 
from \citet{elso85} appropriate for the $s$ types 
\citep{elso88} of the clusters.\\
} \\ 
V\dotfill&      SL~363 & \citet{bern74,bern75,bica96}\\
&       SL~556 & \citet{bica96} \\
&      SL~855 & \citet{mack03}{\Large $^a$}\\
VI\dotfill&      SL~842 & \citet{bica96} \\
&      ESO~121-SC03 & \citet{mate86} \\
VII\dotfill&  NGC~1786 & \citet{vand81} {\bf and} \citet{bica96}
\footnote{\citet{vand81} lists $V = 10.88$ and states that only 
observations corrected for the contribution of a foreground star 
have been considered, but neglects giving a diaphragm. 
\citet{bica96} register both a $V$ mag of 10.88 and a diaphragm 
of 60\arcsec.
}\\
\hline	

\end{tabular}
\end{minipage}
\label{tabphot}
\end{table*}

\begin{table*}
 \centering
 \begin{minipage}{140mm}
  \caption{$\barM_{K_s}$ and ($V - I$) values}
  \begin{tabular}{@{}lcc@{}}
  \hline
  Supercluster & $\barM_{K_s}$& ($V - I$) \\
 \hline
pre\dotfill & -7.70$\pm$0.40&0.34 $\pm$ 0.14  \\
I\dotfill & -8.85$\pm$0.12&  0.61 $\pm$ 0.08\\
II\dotfill & -7.84$\pm$0.28&0.48 $\pm$ 0.10 \\
III\dotfill & -7.45$\pm$0.24  &0.47 $\pm$ 0.06\\
IV\dotfill & -7.51$\pm$0.18&0.54 $\pm$ 0.05\\
V\dotfill & -6.69$\pm$0.20&0.78 $\pm$ 0.11\\
VI\dotfill & -6.21$\pm$0.24&1.02 $\pm$ 0.07\\
VII\dotfill & -4.92$\pm$0.38&1.06 $\pm$ 0.13\\
\hline	
\end{tabular}
\end{minipage}
\label{tabcol}
\end{table*}

\section{Models versus observational data}
\label{results}

\begin{figure*}
\includegraphics[clip=,width=0.45\textwidth]{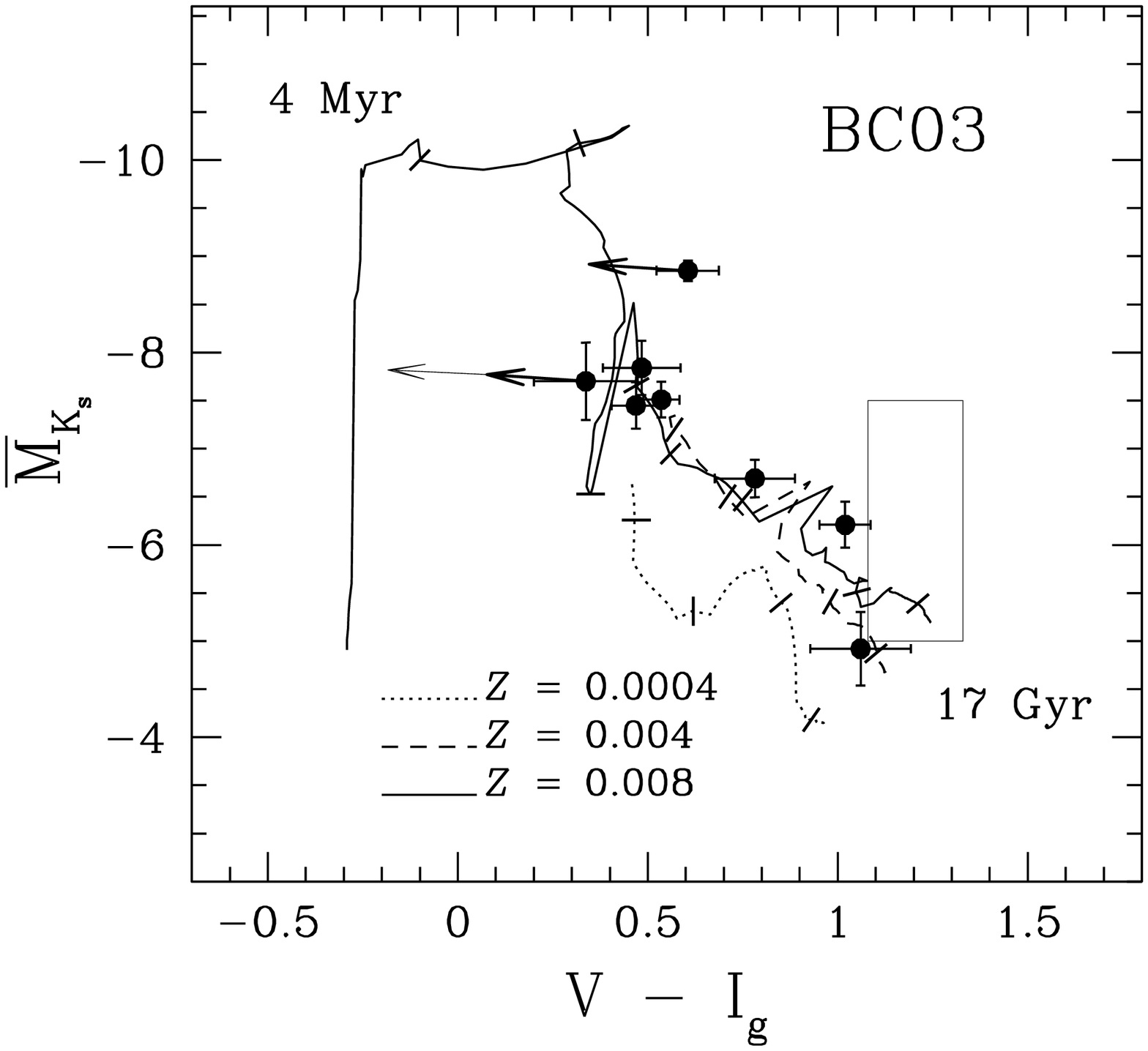}
\includegraphics[clip=,width=0.45\textwidth]{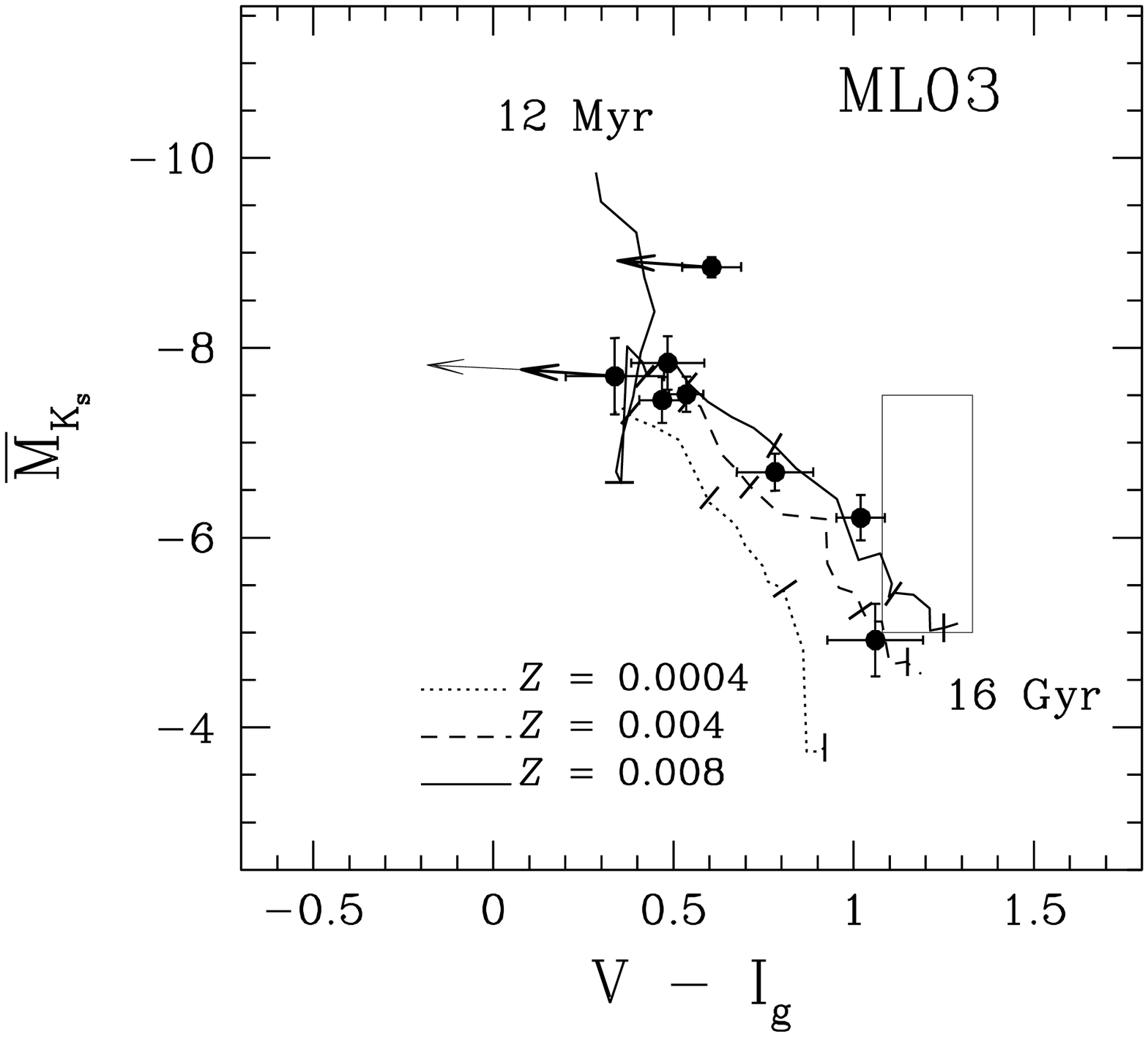}
\caption{Comparison of \barM$_{K_s}$ versus ($V - I_g$) colour 
of Magellanic Cloud superclusters with low-$Z$ models by BC03, {\it left 
panel}), and by ML03 {\it right panel}). {\it Dotted line:} model 
with $Z$ = 0.0004; {\it dashed line:} $Z$ = 0.004; {\it solid line:} 
$Z$ = 0.008. Tickmarks as in Fig.\ \ref{mdl_comp}. The thick arrows 
deredden the Pre-SWB and SWB I superclusters by $E(B - V) = 0.2$ 
above the average of the LMC. The thin arrow dereddens the Pre-SWB 
supercluster by an additional $E(B - V) = 0.15$ (see text). 
The rectangle marks locus of the \citet{liu02} galaxy sample.
}
\label{figres}
\end{figure*}

\begin{figure*}
\includegraphics[clip=,width=0.45\textwidth]{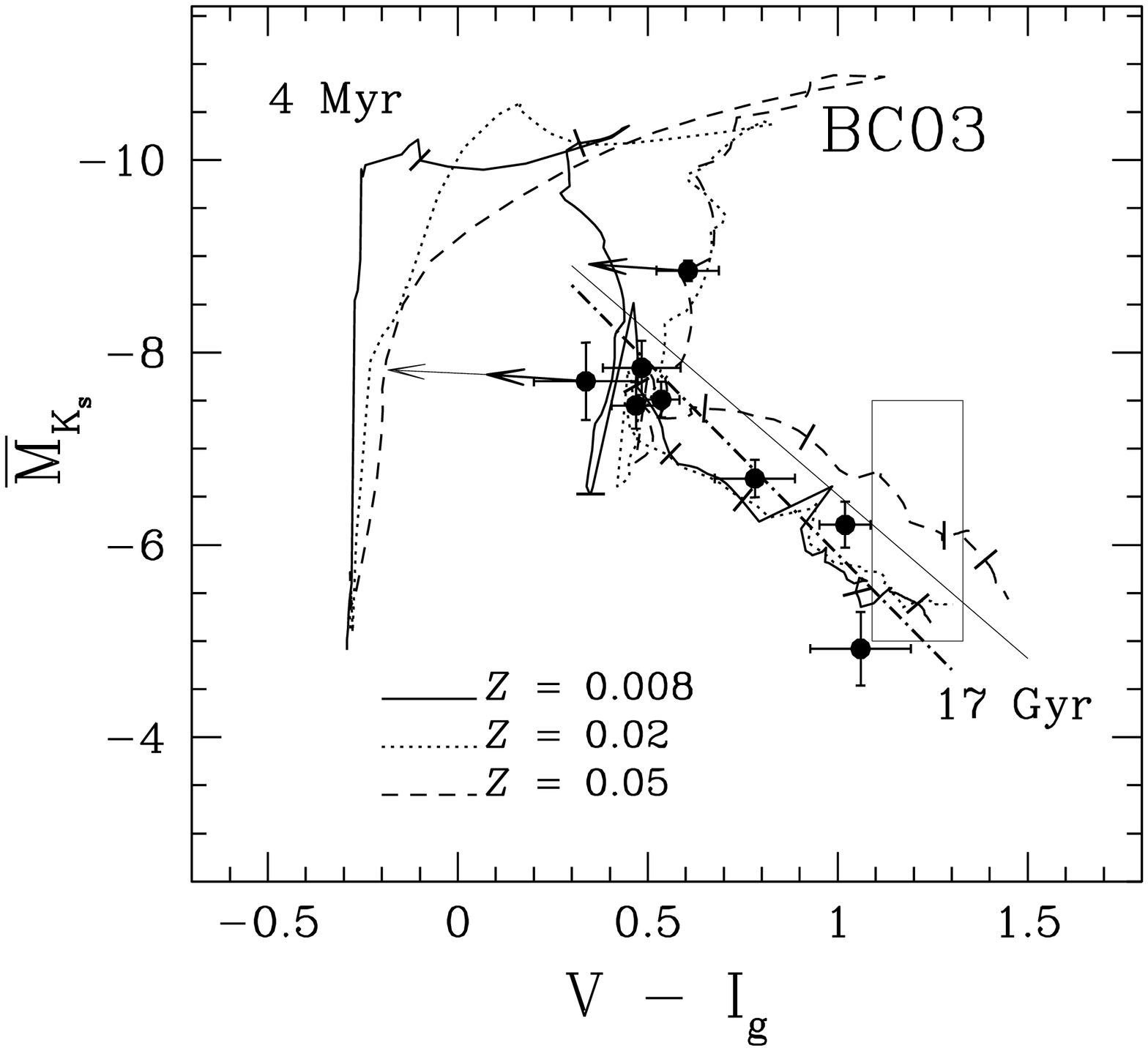}
\includegraphics[clip=,width=0.45\textwidth]{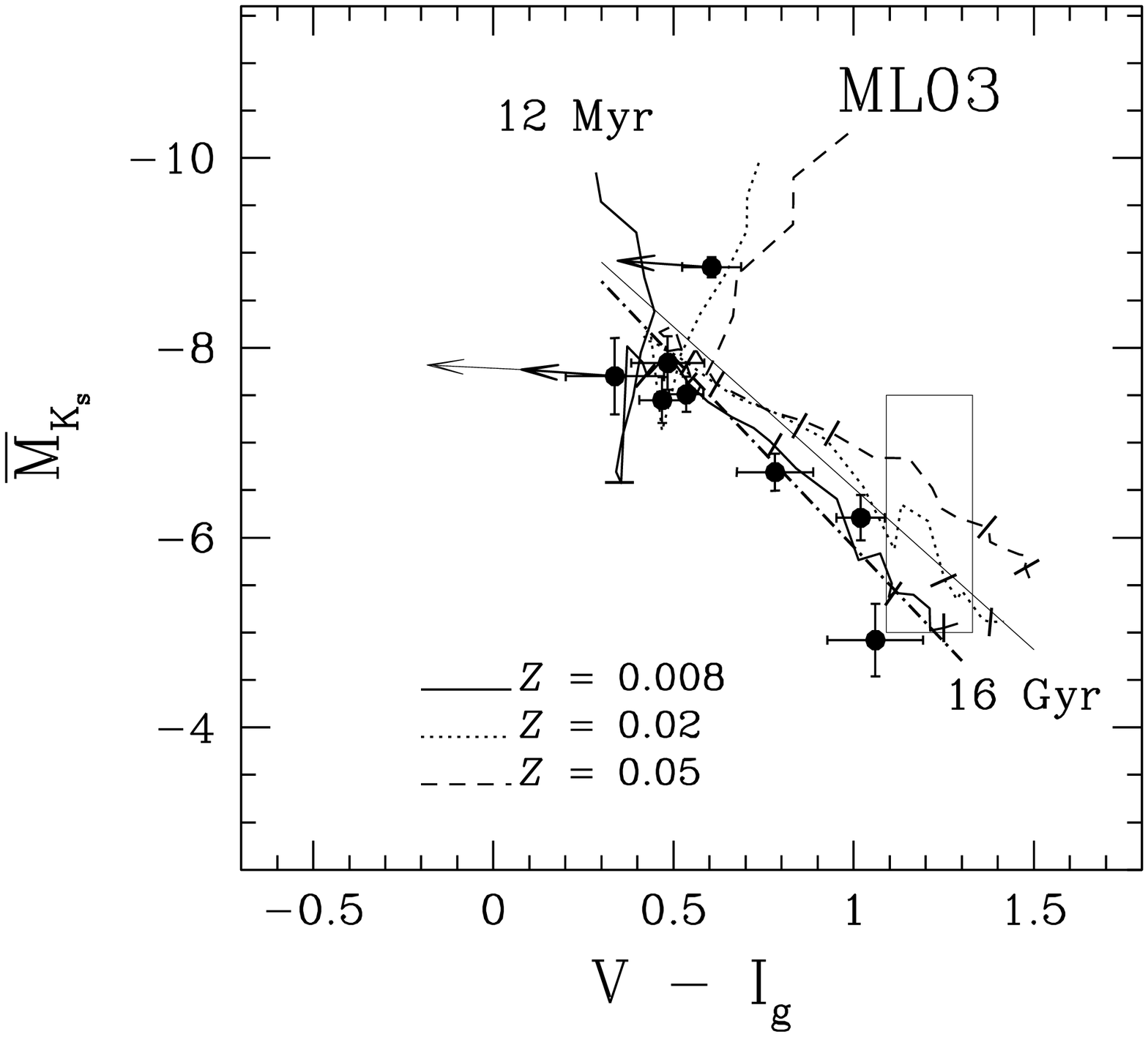}
\caption{Comparison of \barM$_{K_s}$ versus ($V - I_g$) colour of
Magellanic Cloud superclusters with high-$Z$ models by BC03 ({\it left panel}), 
and by ML03 ({\it right panel}). {\it Solid line:} model with 
$Z$ = 0.008; {\it dotted line:} $Z$ = 0.02; {\it dashed line:} $Z$ = 0.05. 
Evolution before $\sim$ 300 Myr is now shown for all models. Symbols 
as in Figures \ref{mdl_comp} and \ref{figres}, except that BC03 model 
with $Z$ = 0.02 is graphed without tickmarks, since it runs so closely 
to the $Z$ = 0.008 model. The straight solid line is best fit to 
\citet{liu02} sample of early-type galaxies and spiral bulges. The 
straight dotted-dashed line is our fit to superclusters I through VI
(see text).}  
\label{figreshighZ}
\end{figure*}

Figures \ref{figres} and \ref{figreshighZ} show the comparison 
of \barM$_{K_s}$ versus ($V - I_g$) colour of MC superclusters 
with both sets of models. BC03 models are presented in the left 
panels, while ML03 ones are displayed on the right.
Fig.\ \ref{figres} presents models with $Z$ = 0.0004, $Z$ = 0.004, 
and $Z$ = 0.008. Even though the superclusters all have $Z$ $\la$ 0.01, 
for completeness Fig.\ \ref{figreshighZ} illustrates the evolution 
of models with solar metallicity and $Z$ = 0.05; the models with 
$Z$ = 0.008 are also shown in Fig.\ \ref{figreshighZ}, for comparison 
purposes. The solid dots represent the MC supercluster data, and 
the rectangle marks the general locus of the \citet{liu02} galaxy 
sample. The observed ($V - I_c$) galaxy colours reported by 
\citeauthor{liu02} were transformed to ($V - I_g$), assuming that 
the mean stellar metallicities of the sample galaxies cover the 
range 0.008 $< Z <$ 0.05, by the following equation:

\begin{equation}
(V - I_g) = (1.050 \pm 0.006)(V - I_c) - (0.005 \pm 0.002)
\label{transeq}
\end{equation}
 
Focusing first on the data and on Fig.\ \ref{figres}, we notice 
that the youngest Pre-SWB and SWB I superclusters, respectively 
at ($V - I$) = 0.34 and ($V - I$) = 0.61, are observed with 
significantly redder optical colours than predicted by the models 
with $Z$ = 0.008, the ones closest to their metallicity of $Z$ = 0.01 
\citep{cohe82}. Given their young ages, though, this is not 
surprising; \citet{char00}  offer the prescription that populations 
younger than $\sim$ 10$^7$ years suffer from about three times more 
reddening than later in their lifetimes.\footnote{Exactly the same
relation between age and reddening was found observationally 
by \citet{vand68} for the MC clusters.} Concordantly, for example, 
\citet{greb00} have measured a total colour excess (including 
foreground extinction) $E(B - V) =$ 0.28$\pm$0.05 for Hodge 301, 
a relatively old cluster in 30 Dor for which these authors also 
derive an age of 20--25 Myr. This is the mean age of the star 
clusters that compose the supercluster type I (Paper I), 
if one adopts the age calibration by \citet{elso85}. Interestingly, 
if one dereddens the SWB I supercluster by an extra 0.20 mag in 
$E(B - V)$, or the difference between the measurement by 
\citet{greb00} and the average 0.08 mag that was adopted for the 
LMC \citep{schl98}, the data point falls exactly on the models 
with $Z$ = 0.008. We follow a similar procedure for the Pre-SWB 
supercluster.
Both \citet{park93} and \citet{dick94} have measured an average 
reddening of $E(B - V) \sim$ 0.43 towards the whole of the 30 Dor 
region, while \citet{selm99} have determined a radially dependent 
reddening in the direction of R136 that reaches $E(B - V) \ga$ 0.5 
at the center of the cluster and declines to $E(B - V) \sim$ 0.3 
at $r =$ 1\arcmin.
Dereddening the youngest, Pre-SWB supercluster by $E(B - V) =$ 0.35
above the average of the LMC places it right where the BC03 models 
predict a population younger than 5 Myr should be in Figures 
\ref{figres} and \ref{figreshighZ}. It is possible
that the fluctuation magnitudes derived for the
Pre-SWB and type I superclusters suffer from a certain degree of
crowding \citep{mouh05}. If this is the case, 
the correct values would be fainter than the reported ones,
but would still lie along the same model 
sequence, at ages, respectively, slightly younger and 
slightly older. For the rest of the MC 
supercluster data points we attempt no further correction.

The evolution predicted by the models 
of the $K_s$-band SBF 
magnitudes as a function of the ($V-I$) colour agrees remarkably 
well with the observed sequence defined by the MC superclusters.
Superclusters of SWB types II, 
III, and IV have ages, respectively, 
of 35, 105, and 320 Myr, 
using the SWB class-age transformation of \citet{frog90}.
Their observed loci concur nicely 
with the predicted location of stellar populations with ages between 
a few $\times \sim$ 10 Myr and $\sim$ 300~Myr, in the region at 
$(V-I)\sim\,0.4$ mag and $\barM_{K_s}\sim\,-7.5$ mag. 
SWB types V, VI, and VII correspond to
ages of $\sim$ 1 Gyr, 3 Gyr, and 9 Gyr, respectively.
For older stellar 
populations, the predicted monotonic fading of the $K_s$-band 
SBF magnitudes as the ($V-I$) colour gets redder
agrees with the observed properties of intermediate-age and old MC 
superclusters.

However, one cannot fail to notice that the MC superclusters 
older than $\sim\,3~$Gyr (i.e, with SWB types VI and VII)
are not very well matched by the models. Given their respective  
ages and metallicities,\footnote{For these SWB types,
$ 0.002\la\,Z\,\la\,0.0008$ \citep{frog90}} the type VI supercluster 
mostly seems brighter in \barM$_{K_s}$, while type VII
mainly appears redder in ($V - I$) than the model predictions. 
We remind the reader here that the measured ($V - I$) colours
could be slightly biased to the red as a consequence of 
our ignorance of the exact telescope pointings used
for the acquisition of the $V$ data (see \S~\ref{denis}). 
We have discussed in the same section that the colour distribution  
of the individual SWB VII clusters shows evidence
of slight population differences, although we could also
be seeing the effects of small-number statistics (12 clusters). 

As advanced in \S~\ref{intro}, it is readily apparent in Figures 
\ref{figres} and \ref{figreshighZ} that the MC superclusters have 
greatly increased the range in \barM$_{K_s}$ versus ($V - I$) that 
is now accessible for exploration. 
We have plotted in both panels of Fig.~\ref{figreshighZ}, as 
a solid straight line, the linear 
fit found by \citet{liu02} to their early-type galaxy and spiral 
bulge data set. After transforming ($V - I_c$) to ($V - I_g$) with 
eq.\ \ref{transeq}, this relation becomes 

\begin{equation}
\barM_{K_s} = (-5.84\pm0.04) + (3.4\pm0.8) [(V - I_g)_0 - 1.20].
\label{fitliu}
\end{equation}

It is remarkable that, with the exception of 
the extremely young supercluster Pre-SWB, and the old,
metal-poor supercluster SWB class VII, 
the remaining MC superclusters lie roughly along the correlation
found by \citet{liu02}. In fact, a fit to superclusters 
I (corrected for extra extinction as explained in 
\S \ref{results}) through VI yields: 

\begin{equation}
\barM_{K_s} = (-5.1\pm0.4) + (4.0\pm0.6) [(V - I_g)_0 - 1.20].
\label{myfitks}
\end{equation}

This is also shown in Fig.\ \ref{figreshighZ}, as a straight
dotted-dashed line.
Within the errors, the relations for, respectively, MC superclusters and 
galaxies have the same slope,\footnote{
If in fact the fluctuation magnitude of supercluster
type I is affected by crowding and its true value is
a few tenths of a mag dimmer, the match between the
two slopes would be even better.}
at the same time that  
the MC superclusters 
have a systematically lower \barM$_{K_s}$ magnitude, at 
a fixed ($V - I$) colour, than what is predicted 
for early-type galaxies. The observed offset can be explained, 
according to both the BC03 and ML03 sets of models, as a metallicity 
effect. The metallicity of the MC superclusters is roughly (bar 
only SWB class VI) half that of the sample of early-type galaxies 
and spiral bulges. For a given ($V-I$) colour, the metal-poor 
simple stellar populations are on average older than the metal-rich 
ones, having then fainter \barM$_{K_s}$ as a consequence of both 
lower metallicities and older ages.

We refer the reader to Paper I for a detailed discussion 
of the reasons for the monotonic decline of the $K_s$-band SBF 
brightness with age displayed by intermediate-age and old 
populations.
When compared to both sets of models, the result obtained here 
for the MC superclusters backs the interpretation offered in 
\citet{liu02}, that their galaxy sample constitutes a sequence. 
Namely, that objects with brighter SBFs have a more recent latest 
star formation burst (or a more extended star formation history), 
and that the most recent bursts in every object have 
all occurred at roughly constant 
metallicity. This is so because their sample lies on a straight
line, parallel to the age sequence of their single metallicity 
models. Simultaneously, 
the fact that there exists a 
discernable offset between the galaxies and the MC superclusters
confirms that the \barM$_{K_s}$ vs. ($V - I$) plane 
may contribute to 
the decoupling of age and metallicity effects in intermediate-age
and old stellar systems.
Particularly worthy of notice is the observation that the two samples 
offer consistent results, considering that galaxies are composite stellar 
systems, while the superclusters are approximately single stellar
populations. The most likely explanation is that, being mostly 
probes of the brightest stars in a population at a given wavelength, 
SBFs are relatively insensitive to an underlying older population
in composite systems.

It is also interesting to remark
here that, while the BC03 models attribute a higher metallicity 
to NGC~1419 and NGC~1389, the two Fornax objects left out of their fit
(through sigma-clipping) by \citet{liu02}, the 
ML03 model with solar metallicity (Fig.\ \ref{figreshighZ}, right 
panel, dotted line) shows an excursion to brighter SBF magnitudes 
at exactly the right place to ascribe to them the same metallicity 
as to other galaxies that do lie on the linear fit. This is a 
consequence of the longer lifetimes of TP-AGB stars in 
ML03 models. Hence, longer TP-AGB lifetimes could partly
explain the observations of galaxies with 
\barM$_{K_s}$ brighter than predicted, given their ($V - I$) colour. 

Last but not least, we want to mention here that the
largest sample to date of galaxies put together for SBF studies
has been presented by \citet{jens03}. The 65 objects in this
sample, however, were observed with the Near-IR Camera and
Multi-Object Spectrometer (NICMOS) camera 2 (NIC2) on board 
the Hubble Space Telescope ({\sl HST}), using the $F160W$ filter. 
The photometric transformations between the NIC2 and the commonly
used near-IR ground-based filters is particularly difficult, 
for several reasons: the very deep molecular absorption bands in
the star themselves; the significant differences between the
{\sl HST} and the ground-based filters; and the fact that 
telluric absorption features are very deep in ground-based
observations but completely absent in data obtained from space
\citep{step00}. 
Dealing with fluctuation magnitudes 
has an added degree of complexity, and conflicting statements 
can be found in the literature with regard to what colour 
to use to determine the correct transformation 
coefficients [i.e., the fluctuation colour of the population
\citep{buzz93,blak01},
or the mean of the integrated and the fluctuation colours
\citep{tonr97}]. With all these caveats, we have transformed
the relation between \barM$_{F160W}$ and ($V - I_c$), derived 
by \citet{jens03} from the 47 galaxies in their sample that
show no signs of dust in the NIC2 field of view, 
into one between \barM$_{H}$ and ($V - I_g$). To this end,
we have relied on eq.\ \ref{transeq}, and on the transformation 
between \barM$_H$ and \barM$_{F160W}$ obtained  
in eq.\ 2 of Paper I.\footnote{
\barM$_H$ = \barM$_{F160W} +$ (0.08 $ \pm$  0.07) $ -$  (0.24 $ \pm$  0.05) ($ J - K_s +$  0.48).}
For ($J - K_s$), we have substituted 0.92 $\pm$ 0.06, which
is the average integrated colour of the mentioned 47 galaxies, as 
obtained from the 2MASS Extended Source Catalog (XSC), 
via the GATOR catalog query page. This is the result:

\begin{equation}
\barM_H = (-5.12\pm0.09) + (4.9\pm0.5) [(V - I_g)_0 - 1.21].
\label{fitjensen}
\end{equation}

For the MC clusters, using the values for \barM$_H$ derived 
in Paper I and in \citet{gonz05}, with the fluctuation magnitude
for the supercluster type I once again corrected for extra
extinction as in \S \ref{results}, the relation reads:

\begin{equation}
\barM_H = (-4.0\pm0.5) + (4.8\pm0.7) [(V - I_g)_0 - 1.21].
\label{myfith}
\end{equation}

Even if tentative, since the near-IR observations of
galaxies and star clusters have not been 
performed through the same filter, the conclusion is the 
same as the one drawn from the $K_s$ fluctuations: 
the slope for early-type galaxies and MC superclusters is
the same within the errors; the MC clusters 
have fainter fluctuation magnitudes at a given 
($V - I_g$) colour, as a result of their lower metallicity.
It would certainly be worthwhile to obtain the data 
needed to perform a fairer comparison of galaxies and
MC star clusters in the $H$ observing window in the future.
Table 4 lists the coefficients of our fits 
to the MC supercluster data; 
the reduced chi--square ($\widetilde{\chi}^2$) and the rms
of the points (in magnitudes) after the fits are also included.

\begin{table*} 
 \centering
 \begin{minipage}{140mm}
  \caption{Fluctuation absolute magnitude vs.\ 
colour}\footnotetext{Fits of the form
$\barM = a + b\ [\ (V - I_g)_0\ -\ {\rm reference\ colour} ]$; 
the number of objects used
for the fit is tabulated as $N$. The resulting rms of the points
(in magnitudes) after the fit and the
reduced chi-square ($\widetilde{\chi}^2$) are also listed.} 
  \begin{tabular}{@{}lccccc@{}}
  \hline
\barM &$a$&$b$&$N$&rms&$\widetilde{\chi}^2$ \\
 \hline
\barM$_{K_s}$ & -5.1 $\pm$ 0.4 & 4.0 $\pm$ 0.6& 6 & 0.85 & 1.02 \\
\barM$_H$ & -4.0 $\pm$ 0.5 & 4.8 $\pm$ 0.7& 6 & 1.04 & 1.28 \\
\hline  
\end{tabular}
\end{minipage}
\label{tabfits}
\end{table*}

\section{Summary \& Conclusions}
\label{concl}

In this paper we have presented the relation between absolute 
$K_s$-band SBF magnitude and ($V-I$) 
integrated colour for more than 180 MC star clusters. The newly reported results
extend to fluctuation magnitudes \barM$_{K_s}$ $\sim$ -9 and optical colour 
($V-I$) $\sim$ -0.4 the linear relation observed already for early-type 
galaxies and spiral galaxy bulges in a more limited 
range of both fluctuation magnitudes and colour
[-5 $\ga$ \barM$_{K_s} \ga$ -7, 1.05 $\la$ ($V-I$) $\la$ 1.25]. 

This empirical relation has been 
compared to the predicted evolution of single age, single 
metallicity stellar population properties in the \barM$_{K_s}$ vs. 
($V - I$) diagram, based on both \citet{bruz03} and 
\citet{mouh03} isochrones. The predicted
evolution and the observed sequence agree quite well over the
ranges of \barM$_{K_s}$ magnitude and ($V - I$) colour 
covered by the data
[-5 $\ga$ \barM$_{K_s} \ga$ -9, -0.4 $\la$ ($V-I$) $\la$ 1.25].
With the exception of the extremely young supercluster Pre-SWB,
and possibly the old and metal poor supercluster class VII, the 
remaining superclusters lie on the linear correlation
already found by \citet{liu02}. Such correlation 
is observed in the ranges 
-5 $\ga$ \barM$_{K_s} \ga$ -9, 0.3 $\la$ ($V-I$) $\la$ 1.25.
The existence of
a linear correlation implies that star clusters, early-type 
galaxies and spiral bulges represent 
an age sequence, where younger stellar populations display
brighter $K_s$-band SBF magnitudes and bluer ($V - I$) colour.
At the same time, the discernable offset between galaxies
and MC star clusters confirms that the \barM$_{K_s}$ vs. ($V - I$) plane
may contribute to distinguish the effects of age and 
metallicity in intermediate-age and old stellar systems.
One other suggestive result that emerges from 
the comparison between the galaxy sample and the 
models is that longer lifetimes of TP-AGB stars
might partly explain galaxies with 
near-IR SBFs that are brighter than anticipated,
given their ($V - I$) colour.
A preliminary comparison between the $H$ 2MASS data of the
MC star clusters and the sample of 47 early-type galaxies and
spiral bulges observed by \citet{jens03} through the
$F160W$ filter leads to the same basic conclusions:
galaxies and star clusters lie along correlations with the same slope,
and there is a slight offset between the star cluster
sample and the galaxies, caused by their different metallicities.

We have found that results from star clusters in 
the MC (i.e., single stellar populations in 
Local Group irregular and consequently relatively 
metal-poor galaxies) agree with those of spiral bulges and 
early-type galaxies (i.e., composite systems with higher
metallicities than the MC), 
located not only in the Local Group, but also in mildly dense 
clusters of galaxies like Fornax and Virgo. The implication is
that the relationship between \barM$_{K_s}$ and ($V - I$) might 
be a fairly robust tool, rather insensitive to environment, at
least in the local universe, for
the study of ages and metallicities of unresolved stellar
populations; could provide additional constraints on star formation histories; 
and aid in the calibration of the $K_s$-band SBFs for cosmological distance
measurements.
In this regard --the determination of cosmological distances--, 
the sensitivity of near-IR SBFs to stellar ages and metallicities, 
and perhaps to the details of AGB evolution
in intermediate-age populations, is a potential 
caveat to bear in mind, albeit not necessarily  
in the local universe or in view of today's 
observing capabilities. For example, \citet{ferre99} 
have analised early-type galaxies
in Coma and 17 clusters at 0.3 $\la z \la$ 0.9; their 
results imply that  
only galaxies smaller than 0.5 $L_*$, 
at redshifts $z \ga$ 0.5, can be expected to harbour populations 
with metallicities below solar.


\section*{Acknowledgments}

We thank the whole DENIS Team, especially G.\ Simon and its PI, N.\ Epchtein, 
for making available de DENIS data. The DENIS project is supported,
in France by the Institut National des Sciences de l'Univers, the 
Education Ministry and the Centre National de la Recherche Scientifique,
in Germany by the State of Baden W\"urtemberg, in Spain by the 
DGICYT, in Italy by the Consiglio Nazionale delle Ricerche, in Austria
by the Fonds zur F\"orderung der wissenschaftlichen Forschung and 
the Bundesministerium f\"ur Wissenschaft und Forschung.
R.A.G.\ and M.A.\ acknowledge 
L.\ Carigi and A.\ Bressan for their very useful 
comments on the manuscript. We thank the referee, Joseph B.\ Jensen,  
for his careful reading of the paper;  
we are grateful for his suggestions.

\vspace*{1cm}

\appendix

\setcounter{figure}{3}

\begin{figure*}
\includegraphics[clip=,angle=-90,width=0.65\textwidth]{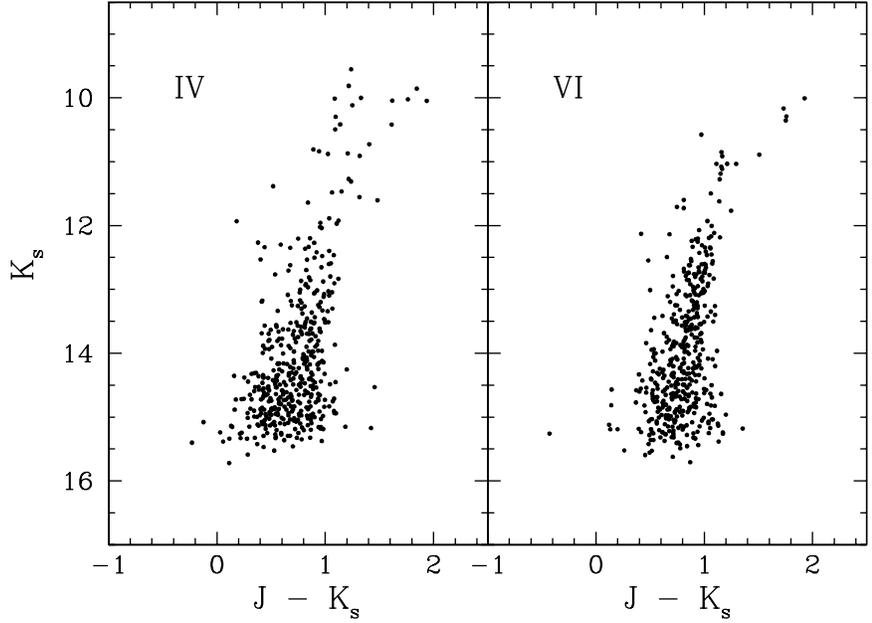}
\caption{Colour--magnitude diagrams of MC superclusters
types SWB IV ({\it left panel}) and SWB VI {\it right panel}).
Stars within 60 arcsec from the center (at the distance of the LMC).
Average photometric errors are 0.04 mag
in brightness and
0.02 mag in color for sources with
$K_s \leq$ 13; 0.06 and 0.03 mag
for stars with 13 $< K_s \leq$ 14; and 0.13 and 0.07 mag
(about the size of the dots) for sources with 14 $< K_s \leq$ 15.
\label{cmds}
}
\end{figure*}

\begin{figure*}
\includegraphics[clip=,width=0.45\textwidth]{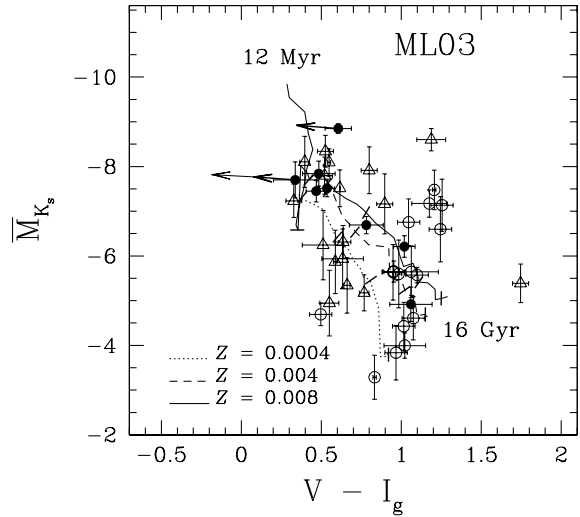}
\caption{\barM$_{K_s}$ versus ($V - I_g$) colour
of individual clusters classes SWB IV ({\it open triangles})
and SWB VI ({\it open circles}); all other symbols as in
right panel of Fig.\ \ref{figres}.
\label{stoch}
}
\end{figure*}

It is beyond the scope of the present paper to investigate
in depth the role of stochastic effects in the integrated
properties of stellar populations. However, for illustration
purposes, we show in Fig.\ \ref{cmds} the colour-magnitude
diagrams (CMDs) of superclusters type IV (left panel) and VI
(right panel). Fig.\ \ref{stoch}  
is a copy of the right panel of Fig.\ \ref{figres},
this time indicating the locations of individual clusters
types IV (open triangles) and VI (open circles) in the
\barM$_{K_s}$ vs. ($V - I$) plane. There are several
facts worth noticing. Firstly, the colour
magnitude diagrams, narrow at the top and wide at the
bottom, resemble more those of SSPs with ``normal"
photometric errors than CMDs of composite populations
(Bressan 2005, private communication). This notwithstanding,
and even when the relative offset between
the general loci of the two SWB classes
is clearly discernible,
single clusters of each SWB type
show a large scatter,\footnote{On the basis of this
plot alone, it could
be argued that perhaps SL~663 , at ($V - I$) $\sim$ 1.7, does
not belong in class SWB IV. Removing it leaves the fluctuation
magnitude and integrated colour of supercluster type IV the
same, within the quoted errors.}
especially in fluctuation magnitude, which is determined
by many fewer stars than the integrated colour. Taken together,
the relatively narrow RGB and AGB, and the very large
scatter of individual clusters in the \barM$_{K_s}$ vs.
($V - I$) plane support the interpretation that
stochastic effects might be more important than
actual population variations in determining the integrated
properties of single clusters within each SWB class.
If this is the case, the construction of superclusters
is an appropriate strategy to simulate more massive
SSPs.

\bsp

\label{lastpage}

\end{document}